\documentclass[11pt]{article}
\usepackage{amsmath,amssymb,color,epsfig,cite}
\usepackage{graphicx}
\usepackage{subfigure}
\usepackage{setspace}

\usepackage{amsfonts}

\newcommand{\hoch}[1]{$\, ^{#1}$}

\makeatletter
\@addtoreset{equation}{section}
\makeatother

\newcommand{\be}{\begin{equation}}
\newcommand{\ee}{\end{equation}}
\newcommand{\bea}{\setlength\arraycolsep{2pt} \begin{eqnarray}}
\newcommand{\eea}{\end{eqnarray}}
\newcommand{\nn}{\nonumber}

\def\ft#1#2{{\textstyle{\frac{\scriptstyle #1}{\scriptstyle #2} } }}
\def\fft#1#2{{\frac{#1}{#2}}}

\def\0{{\sst{(0)}}}
\def\1{{\sst{(1)}}}
\def\2{{\sst{(2)}}}
\def\3{{\sst{(3)}}}
\def\4{{\sst{(4)}}}
\def\5{{\sst{(5)}}}
\def\6{{\sst{(6)}}}
\def\7{{\sst{(7)}}}
\def\8{{\sst{(8)}}}
\def\9{{\sst{(9)}}}

\def\sst#1{{\scriptscriptstyle #1}}

\begin{document}



\begin{center}
{\large {\bf Static Equilibria of Charged Particles Around Charged Black Holes:\\
Chaos Bound and Its Violations}}

\vspace{10pt}

Qing-Qing Zhao\hoch{\dagger}, Yue-Zhou Li\hoch{\ddagger} and H. L\"u\hoch{*}

\vspace{15pt}

{\it Center for Joint Quantum Studies, Tianjin University, Tianjin 300350, China}

\vspace{40pt}

\underline{ABSTRACT}
\end{center}

We study the static equilibrium of a charged massive particle around a charged black hole, balanced by the Lorentz force. For a given black hole, the equilibrium surface is determined by the charge/mass ratio of the particle. By investigating a large class of charged black holes, we find that the equilibria can be stable, marginal or unstable.  We focus on the unstable equilibria which signal chaotic motions and we obtain the corresponding Lyapunov exponents $\lambda$.  We find that although $\lambda$ approaches universally the horizon surface gravity $\kappa$ when the equilibria are close to the horizon, the proposed chaotic motion bound $\lambda<\kappa$ is satisfied only by some specific black holes including the RN and RN-AdS black holes. The bound can be violated by a large number of black holes including the RN-dS black holes or black holes in Einstein-Maxwell-Dilaton, Einstein-Born-Infeld and Einstein-Gauss-Bonnet-Maxwell gravities.  We find that unstable equilibria can even exist in extremal black holes, implying that the ratio $\lambda/\kappa$ can be arbitrarily large for sufficiently small $\kappa$.  Our investigation does suggest a universal bound for sufficiently large $\kappa$, namely $\lambda/\kappa <{\cal C}$ for some order-one constant ${\cal C}$.

\vfill {\footnotesize \hoch{\dagger}zhaoqq@tju.edu.cn\ \ \ \hoch{\dag}liyuezhou@tju.edu.cn\ \ \ \hoch{*}mrhonglu@gmail.com\ \ \ }

\pagebreak

\tableofcontents
\addtocontents{toc}{\protect\setcounter{tocdepth}{2}}


\newpage

\section{Introduction}
\label{sec:intro}

The Newton's gravity between two charged particles can precisely balance the electrostatic Coulomb force for some critical charge/mass ratio. The equilibrium is independent of the separation distance of the particles.  In Einstein-Maxwell theory, these charged particles with the critical charge/mass ratio become asymptotically-flat extremal Reissner-Nordstr\"om (RN) black holes. Despite the high nonlinearity of the theory, the no-force condition between the extremal RN black holes persists and the spacetime geometry can be described by a harmonic function in three dimensional Euclidean space.  The no-force condition is related to the preserved supersymmetry of the spacetime configuration in the context of supergravities.

    The situation becomes much more complicated for non-extremal black holes where the charge/mass ratio is smaller than the critical value. In order to avoid the difficulties associated with the full nonlinearity of Einstein's equation of motion, one typically simplifies the problem by considering a test particle around the black hole. Intriguingly, a test particle can exhibit chaotic motion and the occurrence in axisymmetric spacetimes was studied in \cite{Sota:1995ms}.  It turns out that the unstable orbit signals the existence of the chaotic motion. For a test particle moving around the Schwarzschild black hole, an unstable extremum can exist provided that the particle spins sufficiently fast \cite{Suzuki:1996gm}. It was shown  that null particles can also exhibit chaos around the black holes \cite{Cardoso:2008bp,Dalui:2018qqv}. Considering the fact that black holes are the most fundamental objects in Einstein gravity, it is of great interest to study the universal properties of such chaotic motions of the surrounding test particles.

Recently particle motions moving around the most general static black holes were studied in \cite{Hashimoto:2016dfz}, where external forces such as the static electric force were considered. By focusing on the near-horizon geometry, one can deduce that there is a static unstable equilibrium and the perturbative motion restricted to the radial direction takes a universal form
\be
\epsilon \sim e^{\pm\kappa\, t}\,,
\ee
where $\kappa$ is the surface gravity on the horizon.  It can be argued that this unstable equilibrium implies chaotic motions for general perturbations and furthermore it was conjectured in \cite{Hashimoto:2016dfz} that there is a universal bound for the Lyapunov exponent of chaotic motions caused by the black holes, namely
\be
\lambda\le \kappa\,.\label{unicmb}
\ee
Even though this is a single-particle system, the result appears to agree with the chaos bound proposed in \cite{Maldacena:2015waa} for thermal quantum systems with a large number of degrees of freedom.

We would like to be cautious with the arguments for two reasons.  The first is that a single particle typically does not have to obey constraints derived from a many-body system. The second is that there is no decoupling limit such that the near-horizon geometry of a non-extremal black hole can be a solution on its own.  The sub-leading terms in the near-horizon expansion may have nontrivial contributions.  By including these sub-leading terms, we find that the Lyapunov exponent of the unstable equilibrium at $r_0$ close to the horizon $r_+$ is given by
\be
\lambda^2 = \kappa^2 + \gamma\, (r_0-r_+) + {\cal O}\big((r_0-r_+)^2\big)\,.
\ee
It can be easily demonstrated that no massive particle can have an equilibrium on the horizon of a static non-extremal black hole; therefore, the equality in (\ref{unicmb}) cannot be reached. However, $r_0$ can be arbitrarily close to $r_+$ provided with sufficiently large charge/mass ratio.  The necessary condition that the chaotic motion bound (\ref{unicmb}) is true is that the parameter $\gamma$ is negative.  With explicit examples, we can demonstrate that $\gamma$ can be both negative and positive.

     Even for black holes with negative $\gamma$, in which case there is a local chaotic motion bound (\ref{unicmb}) in the vicinity of the black hole horizon. It is still of interest to investigate whether this bound is globally satisfied by all unstable particle equilibria around the black hole, including regions distance away from the horizon.  We thus study the static equilibria of charged particles around static charged black holes in the whole region rather than only in the vicinity of the black hole horizon. We find that the equilibria can be stable, unstable and/or marginal, depending on the detailed properties of a specific black hole.

The paper is organized as follows.  In section \ref{sec:setup}, we give the general formalism of computing the static equilibrium of charged test particles around charged black holes, assuming that the external force involved is only the Lorentz force in curved spacetime.  We then study the properties of such an equilibrium near the horizon and find that the ``universal property'' can be violated by the sub-leading terms of the metric functions of the near-horizon geometry.  In section \ref{sec:RN}, we study all the static equilibria of RN black holes that are asymptotic to Minkowski, anti-de Sitter (AdS) and de Sitter (dS) spacetimes. We find that the chaos bound (\ref{unicmb}) is satisfied globally for the first two cases, but can be violated in RN-dS black holes by the equilibria some appropriate distance away from the horizon. In section \ref{sec:EMD}, we study asymptotically-flat charged black holes in supergravity-inspired Einstein-Maxwell-Dilaton theories and we find that the bound (\ref{unicmb}) can be violated even locally in the vicinity of the horizon, when the spacetime dimensions are not four.  In section \ref{sec:more}, we study these properties with further charged black hole examples in Einstein-Born-Infeld and Einstein-Gauss-Bonnet-Maxwell gravities.  We conclude the paper in section \ref{sec:con}.

\section{The general set up}
\label{sec:setup}

In this paper, we consider Einstein gravity or its covariant higher-order curvature generalizations, in general $D$ dimensions, coupled to the Maxwell field $A=A_\mu dx^\mu$ and other matter fields, including the cosmological constant, minimally or non-minimally.  We assume that the theories admit the static solutions with the ansatz
\be
ds_{D}^2 = - h(r) dt^2 + \fft{dr^2}{f(r)} + \rho(r)^2 d\Omega_{D-2,k}^2\,,\qquad A= \psi(r) dt\,,\qquad \cdots\,,\label{generalmetric}
\ee
where the ellipses denote any other matter fields that are involved in the solutions, but not relevant for our discussion.  The metric $d\Omega_{D-2,k}^2$ is Einstein with $\widetilde R_{ij} = (D-3)k \tilde g_{ij}$.  Without loss of generality, we can take $k=-1,0,1$, corresponding to hyperbolic, Euclidean or spherical spaces respectively when the metric $d\Omega_{k}^2$ is maximally symmetric.  For asymptotic flat Minkowski spacetimes, we must have $k=1$; for asymptotic AdS or dS spacetimes, we can have $k=-1,0,1$.  Without loss of generality, one can make a coordinate gauge choice $\rho=r$; however, in many explicit black hole examples, it is more convenient or even necessary to make a specific coordinate choice $\rho(r)$ such that the solutions may become analytical.

We further assume that the metric (\ref{generalmetric}) describes a black hole.  In other words, there exists an event horizon (or outer horizon) at $r=r_+$ such that $h(r_+)=0=f(r_+)$.  (In this paper, when we refer to a horizon without an adjective, it always means the event horizon.) The surface gravity on the horizon is then given by
\be
\kappa = \ft12 \sqrt{h'f'}\Big|_{r=r_+}\,,
\ee
where a prime denotes a derivative with respect to $r$.  The surface gravity vanishes for the extremal black holes where $h$ and $f$ have double zeros at $r=r_+$.  We now introduce a test particle charged under the Maxwell field $A$, with mass and charge $(m,e)$.  The motion of this particle caused by the charged black hole is governed by the action
\be
S=-m \int d\tau \left(\sqrt{-g_{\mu\nu} \fft{dx^\mu}{d\tau} \fft{d x^\nu}{d\tau}} + \fft{e}{m} A_\mu \fft{d x^\mu}{d\tau}\right)\,,\label{particleaction}
\ee
where $\tau$ is a certain affine parameter, and the second term in the bracket gives rise to the Lorentz force in curved spacetime.  We are interested in static equilibria of the particle around the black hole; therefore, we focus on the radial motion with no angular momentum.  The relevant action can be written as
\be
S=m \int dt L\,,\qquad L= - \sqrt{h(r) - \fft{\dot r^2}{f(r)}} - \fft{e}{m} \psi(r)\,,
\ee
where the radial variable $r$ is now a function of the asymptotic physical time $t$ and a dot is a derivative with respect to $t$.  We assume that for some appropriate charge/mass ratio $e/m$, there exists an equilibrium hypersurface $r=r_0$.  The purpose of this paper is to examine the stability of this equilibrium.  For small perturbation restricted in the radial direction, with $|\dot r| \ll 1$, the effective Lagrangian is given by
\be
L=\fft{\dot r^2}{2\sqrt{h(r)}\,f(r)} - V_{\rm eff}(r)\,,\qquad V_{\rm eff} = \sqrt{h(r)} + \fft{e}{m} \psi(r)\,.
\ee
Thus for the equilibrium position located at $r_0\ge r_+$, the charge/mass ratio must satisfy
\be
\fft{e}{m} = -\fft{(\sqrt{h})'}{\psi'}\Big|_{r=r_0}\,.\label{emgen}
\ee
It is reasonable to assume that $\psi'$ is regular and non-vanishing on the horizon.  For non-extremal black holes where $h(r_+)$ is a single zero, we have
\be
\Big|\fft{e}{m}\Big|\rightarrow \infty\,,\qquad \hbox{as}\qquad r_0\rightarrow r_+\,.
\ee
Thus no massive particle can have a static equilibrium on the horizon.  For extremal black holes where $h(r_+)$ is a double zero, the situation is different, since $(\sqrt{h})'$ is finite at $r=r_+$, and a massive charged particle with an appropriate charge/mass ratio can have an equilibrium on  the horizon.

The purpose of this paper is to examine the stability of these equilibrium hypersurfaces at $r=r_0$.
For the linear radial perturbation in the vicinity of $r_0$ with $r(t)=r_0 + \epsilon(t)$, the effective Lagrangian now becomes
\be
L=\fft{1}{2\sqrt{h(r_0)}f(r_0)} \Big(\dot \epsilon^2 + \lambda^2 \epsilon^2\Big) + {\cal O}(\epsilon^3) \,,
\ee
where
\be
\lambda^2 =\sqrt{h} f \Big(\fft{\psi''}{\psi'} (\sqrt{h})'-(\sqrt{h})''\Big)\Big|_{r=r_0}\,.
\label{lambdagen}
\ee
The characteristics of the equilibrium is specified by the sign choice of $\lambda^2$.  Specifically, we have
\be
\left\{
  \begin{array}{ll}
    \lambda^2>0: & \hbox{unstable, with}\qquad \epsilon\sim e^{\lambda t}\,; \\
    \lambda^2=0: & \hbox{marginal;} \\
    \lambda^2<0: & \hbox{stable, with}\qquad \epsilon \sim \cos(\sqrt{-\lambda^2}\, t)\,.
  \end{array}
\right.
\ee
In particular, for the unstable case, although the perturbation restricted to the radial direction can be solved exactly, the general perturbative motion becomes chaotic and the parameter $\lambda$ is the upper bound for the Lyapunov exponent \cite{Sota:1995ms}.  In this paper, we shall not be so pedantic and simply refer to $\lambda$ as the Lyapunov exponent of the local chaotic motion in the vicinity of the unstable equilibrium. (Note that for neutral massive particles, the existence of a static equilibrium requires $h'(r_0)=0$ at certain $r_0$. Such case can arise in asymptotically-dS solutions where there exists an additional cosmological horizon.  Explicit examples will be given in section \ref{sec:RNdS}.)

This behavior was investigated in literature where the equilibrium near the horizon was considered \cite{Hashimoto:2016dfz}.  One may focus on the near-horizon geometry which is specified by the metric functions whose Taylor expansions on the horizon are
\be
f(r)=f_1 (r-r_+) + \cdots\,,\quad h(r)=h_1 (r-r_+) + \cdots\,,\quad \psi(r)=\psi_0+\psi_1 (r-r_+) + \cdots\,,
\label{nearhorizon}
\ee
where $\psi_0$ is pure gauge. If one treats these as the full metric functions, then the equilibrium is located at
\be
r_0=r_+ + \fft{m^2 h_1}{4e^2\psi_1^2}\,.\label{simpler0}
\ee
Perturbing around this $r=r_0$, one finds an universal expression \cite{Hashimoto:2016dfz}
\be
\lambda=\kappa\,.\label{res1}
\ee
However we should be cautious about this ``universal'' result.  It is important to note that for any black hole with non-vanishing surface gravity, there can be no decoupling limit such that the near-horizon geometry can be a solution on its own.  It follows that one cannot always ignore the higher-order terms in the Taylor expansion (\ref{nearhorizon}) even if one is interested only in studying the near-horizon properties.  In particular, the equilibrium (\ref{simpler0}) is valid only for the large $e/m$ ratio; however, there is lacking of a dimensionless parameter to measure this ``largeness''. Nevertheless one could be still tempted to conjecture that (\ref{res1}) provides a universal upper bound for the Lyapunov exponent for chaotic motions outside the black hole horizon, albeit the equality cannot be saturated.

    To address this question, it is informative to include the next order of the near horizon Taylor expansions, namely
\bea
f(r) &=& f_1 (r-r_+) + f_2 (r-r_+)^2 +\cdots\,,\nn\\
h(r) &=& h_1 (r-r_+) + h_2 (r-r_+)^2 + \cdots\,,\nn\\
\psi(r)&=&\psi_0+\psi_1 (r-r_+) + \psi_2 (r-r_+)^2 + \cdots\,.
\label{nearhorizon2}
\eea
At this order, the ratio $e/m$ and the equilibrium $r_0$ is related by
\be
\fft{e}{m} = -\fft{\sqrt{h_1}}{2\psi_1 \sqrt{r-r_+}} +
\fft{-3 h_2 \psi_1 + 4 h_1 \psi_2}{4\sqrt{h_1} \psi_1^2} \sqrt{r-r_+} + \cdots\,.
\ee
The Lyapunov exponent is now modified to become
\be
\lambda^2 = \kappa^2 + \gamma (r_0-r_+) + {\cal O}((r-r_+)^2)\,,\qquad \gamma = \ft14 (f_2 h_1 - f_1 h_2) +4\kappa^2 \fft{\psi_2}{\psi_1}\,.
\ee
Thus we see that the static equilibria sufficiently close to the horizon are all unstable; however,
the earlier near-horizon result (\ref{res1}) could be exactly true only when the equilibrium were located literally on the horizon, which, we have demonstrated, is not possible for any massive particle around such a non-extremal black hole.  The true statement is that $\lambda$ approaches $\kappa$ universally as the equilibrium closes to the horizon, and the horizon surface gravity $\kappa$ may  provide a bound for the Lyapunov exponent for the unstable equilibria in the vicinity of horizon. Whether it is an upper bound or a lower bound depends on the sign choice of $\gamma$:
\be
\left\{
  \begin{array}{ll}
    \gamma>0, &\qquad \hbox{$\kappa$ is locally a lower bound, \it i.e.}\quad \lambda>\kappa\,; \\
    \gamma<0, &\qquad \hbox{$\kappa$ is locally an upper bound, \it i.e.}\quad \lambda <\kappa\,.
  \end{array}
\right.
\ee
The necessary condition for satisfying the universal chaotic motion bound (\ref{unicmb}) is $\gamma<0$.
Since there is no obvious universal energy condition that can enforce the negativity of $\gamma$, the bound (\ref{unicmb}) may likely be violated.

    For a special class of black holes, referred to as special static solutions in \cite{Li:2017ncu}, some more general statements can indeed be made.  These special static metrics are characterised by $h=f$ in the $\rho=r$ coordinate gauge \cite{Li:2017ncu}.  In this case, the electric potential for the minimally coupled Maxwell field is given by
\be
\psi = \fft{q}{r^{D-3}}\,,
\ee
where $q$ is the charge parameter. In this case, we have
\be
\gamma = - \fft{2(D-2)\kappa^2}{r_+} <0\,.
\ee
We see that for these special static black holes there is indeed a ``local'' universal bound (\ref{unicmb}).  We call it a local bound because at this stage we can only be sure that it is valid in the vicinity of the horizon, in other words, for the equilibria hypersurfaces located at $r_0\in (r_+, r_+ + \epsilon)$ for sufficiently small $\epsilon$.  This local analysis says nothing about the Lyapunov exponent associated with the unstable equilibria some distance away from the horizon.

    In the following sections, we study a variety of charged black holes and study the equilibria of charged particles and determine the characteristics of these equilibria.  We focus on finding examples that violate the bound (\ref{unicmb}) in different categories of black holes.

\section{Reissner-Nordstr\"om black holes}
\label{sec:RN}

In this section, we consider Einstein-Maxwell gravity, coupled to a bare cosmological constant $\Lambda_0$:
\be
{\cal L} = \sqrt{-g} (R - \ft14 F^2 -2 \Lambda_0)\,,\label{einmaxlag}
\ee
where $F=dA$ is the field strength.  For simplicity of our presentation, we shall focus our discussion on $D=4$ dimensions and give a brief summary of the results in general dimensions at the end of this section.

\subsection{Asymptotically flat}
\label{sec:RNflat}

In this subsection, we set $\Lambda_0=0$ so that the maximally-symmetric vacuum is the flat Minkowski spacetime.  The asymptotically-flat RN black hole in four dimensions is then given by
\be
h=f= 1 -\fft{2M}{r} + \fft{q^2}{r^2}\,,\qquad \psi = \fft{2q}{r}\,,\qquad \rho=r\,,
\ee
Note that the factor ``2'' in $\psi$ is due the ``1/4'' normalization the kinetic term of the Maxwell field in (\ref{einmaxlag}).  In this normalization, the electric charge is given by
\be
Q_e=\fft{1}{16\pi} \int {*F} = \ft12 q\,.
\ee
This solution belongs to the special static black holes discussed in section \ref{sec:setup} and hence the general statements of the characteristics of the equilibria near the horizon hold true. Nevertheless, it is instructive to study the case in detail so as to uncover the properties of the equilibria away from the horizon.

For sufficiently large mass $M$, there are two horizons, the inner horizon at $r_-$ and outer horizon at $r=r_+$, with $r_-\le r_+$, defined by $f(r_\pm)=0$.  The mass and charge parameters $(M,q)$ can now be expressed as
\be
M=\ft12 (r_+ + r_-)\,,\qquad q=\sqrt{r_+ r_-}\,.
\ee
The surface gravity on the outer horizon is
\be
\kappa=\fft{r_+-r_-}{2r_+^2}\,.
\ee
The requirement that $\kappa\ge 0$ implies that mass and charge must satisfy the inequality
\be
M\ge q=2 Q_e\,,\label{bhcond}
\ee
which is saturated in the extremal limit where the $r_\pm$ coalesce.

It follows from (\ref{emgen}) that the equilibrium $r_0$ is determined by
\be
\fft{e}{m} = \fft{(r_+ + r_-)r_0 - 2 r_+ r_-}{4\sqrt{r_+ r_- (r_0-r_+)(r_0-r_-)}}\,.
\label{emr0forrnflat}
\ee
As was discussed in section \ref{sec:setup}, for $r_+>r_-$, the equilibrium $r_0$ cannot be located on the horizon for any massive particle. For $r_0$ lying in the region $(r_+,\infty)$, we must have
\be
\fft{e}{m}\ge \fft{r_+ + r_-}{4\sqrt{r_+ r_-}} = \fft{M}{2q}=\fft{M}{4Q_e}\ge \ft12\,.
\ee
Thus equilibrium outside of the horizon is only possible for electron-like charged particles that
violate the black hole condition (\ref{bhcond}).  The Lyapunov exponent can be obtained from (\ref{lambdagen}) straightforwardly:
\be
\lambda=\fft{r_+ - r_-}{2r_0^2}\,.
\ee
Since $\lambda$ is always real, the equilibria are all unstable. Furthermore, $\lambda$ satisfies the global upper bound (\ref{unicmb}), since we we have $r_0> r_+$.

When the black hole is extremal, with $r_-=r_+$, (i.e.~$M=2 Q_e$,) it follows from equilibrium condition (\ref{emr0forrnflat}) that we must have $m=2e$, and the equilibria can be at any $r_0$, including on the horizon.  The equilibria are all marginally stable with $\lambda=0$. This is consistent with the well-known result that there is a no-force condition of supersymmetric static black holes that are asymptotically flat.

For the non-extremal RN black hole, outside of the horizon, there exists an equilibrium hypersurface $r=r_0$ for a charged particle, determined by the charge/mass ratio, for the ratio $e/m > 1/2$.  The equilibrium is unstable. The bigger the ratio, the closer is the equilibrium to the horizon and the larger the Lyapunov exponent, but with the upper bound (\ref{unicmb}), which cannot be saturated by any massive particle.

Following the discussions in section \ref{sec:setup}, we expect that there is a local bound of (\ref{unicmb}) at the vicinity of the horizon. It is nontrivial that the bound becomes global for all the equilibria outside the horizon. Is this feature universal? We continue this study by introducing the cosmological constant.

\subsection{Asymptotically AdS}
\label{sec:RNAdS}

We now turn on the cosmological constant $\Lambda_0$.  We take it to be negative $\Lambda_0=-3/\ell^2$.  The maximally-symmetric vacuum is the AdS spacetime of radius $\ell$.  The static and asymptotically-AdS solution is given by
\be
f=\fft{r^2}{\ell^2} + k - \fft{2M}{r} + \fft{q^2}{r^2}\,,\qquad
\psi=\fft{2q}{r}\,,\qquad \rho=r\,.
\ee
Now the topological parameter $k$ can take all $(-1,0,1)$ values.  The solutions in general have two horizons $r_\pm$, and it is convenient to express $(M,q)$ in terms of $(r_-,r_+)$:
\be
M=\fft{(r_+ + r_-) (r_+^2 + r_-^2 + k \ell^2)}{2\ell^2}\,,\qquad
q=\fft{\sqrt{r_+ r_- (r_+^2 + r_-^2 + r_+ r_- + k\ell^2)}}{\ell}\,.
\ee
The surface gravity on the horizon is
\be
\kappa=\frac{\left(r_+-r_-\right) \left(k \ell^2+r_-^2+3 r_+^2+2 r_- r_+\right)}
{2 \ell^2 r_+^2}\,.
\ee
The equilibrium $r_0$ is related to the mass/charge ratio by
\be
\fft{e}{m} =\fft{r_0^2 f'(r_0)}{4q \sqrt{f(r_0)}} \,.
\ee
The corresponding Lyapunov exponent is
\bea
\lambda^2 &=& \frac{-2 r_0^6+3 \left(r_-^2+r_+^2\right) r_0^4-6 r_-^2 r_+^2 r_0^2+r_-^2 r_+^2 \left(r_-^2+r_+^2\right)}{\ell^4 r_0^4}\nn\\
&&+\frac{2 M \left(-3 r_0^4+4 \left(r_-+r_+\right) r_0^3-6 r_- r_+ r_0^2+r_- r_+ \left(r_-^2-r_+ r_-+r_+^2\right)\right)}{\ell^2 \left(r_-+r_+\right) r_0^4}\nn\\
&&+\frac{M^2 \left(r_--r_+\right){}^2}{\left(r_-+r_+\right){}^2 r_0^4}\,.
\eea
Note that for the topological parameter $k=-1$, horizon can exist even for negative mass.  Here we insist that the black hole mass $M$ takes only the positive values.

We first examine the extremal black holes with $r_-=r_+$.  In this case, we have
\bea
\fft{e}{m} &=&\frac{\sqrt{r_+} \left(\ell^2 M+r_+^3+r_0 r_+^2+r_0^2 r_++r_0^3\right)}{2\ell q\sqrt{\ell^2 M+r_+ \left(r_++r_0\right){}^2}}\,,\nn\\
\lambda^2 &=& -\frac{\left(r_0-r_+\right){}^3 \left(\ell^2 M \left(r_++3 r_0\right)+2 r_+ \left(r_++r_0\right){}^3\right)}{\ell^4 r_+ r_0^4}\,.
\eea
Thus we see that for the ratio
\be
\fft{e}{m}=\ft12 \sqrt{1 + \ft{3r_+^4}{\ell^2 q^2}}\,,
\ee
the equilibrium is located on the horizon, for which $\lambda=0$, giving rise to the marginally stable equilibrium.  The concept of no-force condition breaks down for the asymptotically-AdS extremal black holes.  For particles with the larger $e/m$ ratio, the equilibrium $r_0$ is located outside of the horizon and all these equilibria are stable since $\lambda^2<0$.  It is intriguing that charged massive particles can be trapped in these hypersurfaces. This may provide a mechanism of matter condensation.

For the non-extremal black holes with $r_+>r_-$, we have
\bea
r_0\rightarrow r_+: &&\qquad \lambda^2 = \kappa^2 - \fft{4\kappa^2}{r_+} (r_0-r_+) + {\cal O}\big((r_0-r_+)^2\big)\,,\nn\\
r_0\rightarrow \infty: &&\qquad \lambda^2 = - \fft{2r_0^2}{\ell^4} + {\cal O}(1)\,.
\eea
Thus we see, not surprisingly, that $\lambda^2>0$ near the horizon, but it is always negative at asymptotic infinity.  In fact for the non-extremal RN-AdS black holes, there exists $r_0^*>r_+$ where $\lambda=0$.  In the region $(r_+,r_0^*)$, the equilibria are unstable with $\lambda^2>0$.  For $r_0>r_0^*$, we have $\lambda^2<0$ and the equilibria become all stable.

It is of interest to investigate whether the bound (\ref{unicmb}) holds for all the unstable equilibria.
To do so, we define two positive dimensionless parameters $(x,y)$, by
\be
r_0 = (1 + x)r_+\,,\qquad  r_+ = (1 + y)r_-\,.\label{xydef}
\ee
We find that
\bea
\kappa^2 - \lambda^2 &=& \fft{x}{r_-^4 (x+1)^4 (y+1)^4 (y+2)^2}\Big[M^2 \left(x^3+4 x^2+6 x+4\right) y^2\nn\\
&&+\fft{2M}{\ell^2}r_-^3 \left(y^2+3 y+2\right)\Big(x^3 \left(4 y^3+11 y^2+9 y+3\right)+4 x^2 \left(3 y^3+7 y^2+4 y+1\right)\nn\\
&&\qquad\qquad+6 x y \left(2 y^2+4 y+1\right)+4 y^2 (y+2)\Big)\nn\\
&& + \fft{r_-^6 (x+2) (y+1)^2 (y+2)^2}{\ell^4}\Big(2 x^4 (y+1)^4+8 x^3 (y+1)^4 +2 y^2 (y+2)^2\nn\\
&&+x^2 \left(12 y^4+48 y^3+67 y^2+38 y+8\right)+2 x y \left(4 y^3+16 y^2+19 y+6\right)
\Big)
\Big]\,.
\eea
This quantity is positive definite and it approaches zero as $x\rightarrow 0$.  Thus the chaotic motion bound (\ref{unicmb}) is globally satisfied for the RN-AdS black holes.

\subsection{Asymptotically dS}
\label{sec:RNdS}

We now consider positive cosmological constant $\Lambda_0$.  The black hole solution is given by
\be
f=-\ft13 \Lambda_0\, r^2 + 1 - \fft{2M}{r} + \fft{q^2}{r^2}\,,\qquad
\psi=\fft{2q}{r}\,,\qquad \rho=r\,.
\ee
Here, we have chosen $k=1$. In addition to the inner and outer horizons, there exists also the cosmic horizon $r_c$.  The normal spacetime is sandwiched between $r_+$ and $r_c$, where the metric functions $h=f$ are positive. It is helpful to express the constants $M,q$ and $\Lambda_0$ in terms of $r_c> r_+\ge r_->0\,$:
\bea
M &=& \frac{\left(r_-+r_+\right) \left(r_c+r_-\right) \left(r_c+r_+\right)}{2 \left(r_c^2+\left(r_-+r_+\right) r_c+r_-^2+r_+ r_-+r_+^2\right)}\,,\nn\\
q &=& \sqrt{\frac{r_- r_+ r_c \left(r_c+r_-+r_+\right)}{r_- \left(r_c+r_+\right)+r_c^2+r_+ r_c+r_-^2+r_+^2}}\,,\nn\\
\Lambda_0 &=& \frac{3}{r_c^2+\left(r_-+r_+\right) r_c+r_-^2+r_+ r_-+r_+^2}\,.
\eea
The surface gravity on the horizon $r_+$ is given by
\be
\kappa = \frac{\left(r_+-r_-\right) \left(r_c-r_+\right) \left(r_c+r_-+2 r_+\right)}{2 r_+^2 \left(r_c^2 + r_+^2 + r_-^2 + r_- r_c+r_+ r_c+r_+ r_-\right)}\,.
\ee

We now consider the extremal black hole with $r_-=r_+$ and hence $\kappa=0$.  The equilibrium $r_0$ is determined by
\be
\fft{e}{m} = -\fft{r_+^2 \left(r_0-2 r_c\right)+r_+ \left(r_0^2-r_c^2\right)+r_0^3}{
2r_+ \sqrt{r_c \left(r_c+2 r_+\right) \left(r_c-r_0\right) \left(r_c+2 r_++r_0\right)}}\,.
\ee
Since $r_0$ runs from $r_+$ to $r_c$, we find that
\be
\sqrt{\fft{(r_c-r_+) (r_c + 3 r_+)}{4 r_c (r_c + 2 r_+)}} \ge \fft{e}{m}> -\infty\,.
\ee
Note that the equilibrium cannot be located on the cosmic horizon for any massive particle.  The Lyapunov exponent is given by
\bea
\lambda^2 &=& \fft{(r_0-r_+)^3}{r_0^4 (r_c^2 + 2 r_c r_+ + 3 r_+^2)^2} \Big[
(r_0-r_+)(r_0 + r_+)^2\nn\\
&&\qquad+ 2 (3r_0^2 + 4 r_0 r_+ + r_+^2) (r_c - r_0) +
(3r_0 + r_+) (r_c - r_0)^2\Big]\ge 0\,.
\eea
Thus the equilibrium on the horizon is marginally stable with $\lambda=0$, whilst any equilibrium outside the horizon is unstable.  Furthermore, the chaotic motion bound (\ref{unicmb}) is (maximally) violated by all the unstable equilibria since we now have $\kappa=0$ for the extremal black holes. This also implies that $\lambda/\kappa$ can be arbitrarily large for sufficiently small $\kappa$.

The situation becomes more complicated for non-extremal black holes.  The equilibrium $r_0$ is determined by the mass/charge ratio as
\be
\fft{e}{m}=\frac{\left(r_-+r_+\right) r_0 \left(r_c+r_-\right) \left(r_c+r_+\right)-2 r_- r_+ r_c \left(r_c+r_-+r_+\right)-2 r_0^4}{4 \sqrt{r_- r_+ r_c \left(r_0-r_-\right) \left(r_0-r_+\right) \left(r_c-r_0\right) \left(r_c+r_-+r_+\right) \left(r_c+r_-+r_++r_0\right)}}\,.
\ee
Thus there exists an equilibrium for any massive particle.  The bigger the positive $e/m$ ratio, the closer to the event horizon the equilibrium is; the more negative the ratio, it is closer to the cosmological horizon.  The Lyapunov exponent is given by $\lambda^2=X/Y$, with
\bea
X &=&x^2 \Big(2 x^2 (y+1)^2+x \left(5 y^2+9 y+4\right)+y (3 y+4)\Big)^2\cr
&&+(x+1)^2 z^2 \Big(12 x^4 (y+1)^4+16 x^3 (2 y+1) (y+1)^3+6 x^2 y (5 y+4) (y+1)^2\cr
&&\qquad\qquad+6 x y^2 \left(3 y^2+7 y+4\right)+y^2 (3 y+4)^2\Big)\cr
&&+2 x (x+1) z \Big(12 x^4 (y+1)^4+10 x^3 (5 y+4) (y+1)^3+y^2 (3 y+4)^2\cr
&&\qquad\qquad+2 x^2 \left(37 y^2+56 y+16\right) (y+1)^2+3 x y \left(15 y^3+47 y^2+48 y+16\right)\Big)\cr
&&+(x+1)^4 y^2 (y+1)^2 z^4+2 (x+1)^3 y^2 (y+1) z^3 (2 x (y+1)+3 y+4)\,,\nn\\
Y &=& 4 r_-^2 (1 + x)^4 (1 + y)^2 \Big(6 + 4 z + z^2 + x^2 (1 + y)^2 (1 + z)^2
 + y (8 + 7 z + 2 z^2)\nn\\
&& + x (1 + y) (1 + z) (4 + 3 y + 2 z + 2 y z) + y^2 (3 + 3 z + z^2)\Big)^2\,.
\eea
Here $(x,y)$ are defined by (\ref{xydef}) and $z$ is given by
\be
r_c = r_0 (1 + z)\,.
\ee
Since $(x,y,z)$ are all positive values, the quantity $\lambda^2$ must be positive since it is a rational polynomial of $(x,y,z)$ with all positive coefficients. It follows that all the equilibria are unstable.  This is of course consistent with the known fact that a positive cosmological constant cannot balance the matter to give a steady-state universe.  The chaotic bound (\ref{unicmb}) can be violated for sufficiently small surface gravity or sufficiently large cosmological constant.  As a concrete example, we consider $(r_-, r_+, r_c)=(1,2,3)$, corresponding to
\be
(M,q,\Lambda_0)=(\ft65,\ft65, \ft{3}{25})\,,\qquad \kappa=\ft{1}{25}\,,\qquad
f=\fft{(r-1)(r-2)(3-r)(6+r)}{25 r^2}\,.
\ee
The normal spacetime region lies in $r\in (2,3)$. We find
\be
\kappa^2 - \lambda^2 = \fft{2(r_0-2)(3-r_0)(18-5r_0-r_0^2)}{625r_0^2}\,.
\ee
Since the last bracket in the numerator has one native root, and one positive root $r_0^*$:
\be
2<r_0^*=\ft12(\sqrt{97}-5)<3\,,
\ee
It follows that the chaotic bound is violated not near the horizon $r_+$, but in the region $r_0\in (r_0^*,3)$.  In particular, the maximal violation occurs at $r_0=2.74$, which is not near to any horizon.  In fact, we have $\kappa=\lambda$ on both the event and cosmological horizons.

Before ending this subsection, we would like to point out that there is in general an equilibrium hypersurface for neutral massive particle around asymptotically-dS black holes.  It is located at $r_0$ such that $h'(r_0)=0$.  As a concrete example, we consider Schwarzschild-dS black hole with
\be
h=f=-\ft13 \Lambda_0 r^2 + 1 - \fft{2M}{r}\,,\qquad \rho=r\,.
\ee
The black hole as an event horizon $r_+$, as well as a cosmological horizon $r_c>r_+$.  The mass and cosmological constant are related to the two horizons by
\be
M=\fft{r_c r_+ (r_c + r_+)}{2 (r_c^2 + r_+^2 + r_c r_+)}\,,\qquad
\Lambda_0=\fft{3}{r_c^2 + r_+^2 + r_c r_+}\,.
\ee
It is easy to verify that the equilibria is located at
\be
r_0 = \big(\ft{3M}{\Lambda_0}\big)^{\fft13}=\big(\ft12 r_c r_+ (r_c + r_+)\big)^{\fft13}\,,\qquad
\hbox{with}\qquad \lambda^2 = \Lambda_0 - (9 M^2 \Lambda_0^4)^{\fft13}\,.\label{schds}
\ee
Requiring that $r_c>r_+>0$ implies that $\lambda^2>0$ in general and hence the equilibrium is unstable, causing chaotic motion for general perturbation. The Lyapunov index however satisfy the bound (\ref{unicmb}).

\subsection{General dimensions}
\label{sec:genD}

We have so far analysed the RN black holes in four dimensions.  In general $D$ dimensions, the solution is given by ($\rho=r$)
\be
f=-\fft{2\Lambda_0}{(D-1)(D-2)}\, r^2 + k - \fft{2M}{r^{D-3}} + \fft{q^2}{r^{2(D-3)}}\,,\qquad \psi = \sqrt{\fft{2(D-2)}{D-3}}\,\fft{q}{r^{D-3}}\,.\label{genrnsol}
\ee
The qualitative results in general dimensions are the same as those we discussed in $D=4$.  For asymptotic flat solution with $\Lambda_0=0$ and $k=1$, there are in general two horizons $r_+\ge r_-$, and we find that
\be
\kappa = \fft{(D-3) (r_+^{D-3} - r_-^{D-3})}{2 r_+^{D-2}}\,,\qquad
\lambda = \fft{(D-3) (r_+^{D-3} - r_-^{D-3})}{2 r_0^{D-2}}\,.
\ee
In the extremal limit, with $r_-=r_+$, for the particles with the same extremal charge/mass ratio, there is a no-force condition, and the particles are in marginally-stable equilibrium in any space.  For non-extremal black holes, we have real Lyapunov exponent $\lambda$, indicating that the equilibria are all unstable, but the resulting motion satisfies the chaotic bound $\lambda <\kappa$.

For asymptotically AdS black holes, stable equilibria can arise in extremal or near-extremal black holes.  For non-extremal black holes, the equilibria sufficiently far away from the horizon are stable, whilst those close to the horizon are unstable, but satisfy the chaotic motion bound (\ref{unicmb}). For the asymptotically dS RN black holes, all the equilibria are unstable and the chaotic bound can be violated, especially for extremal or near-extremal black holes.

\section{Einstein-Maxwell-Dilaton theory}
\label{sec:EMD}

In the previous section, we examine the properties of the equilibrium hypersurfaces of charged particles around the RN black holes.  For asymptotically flat or AdS black holes, we find that for the unstable equilibria, there is a global upper bound (\ref{unicmb}) for the Lyapunov exponent.  However, this bound is violated for asymptotically dS RN black holes, it is of interest to look for examples of asymptotically-flat or AdS black holes that may also violate this bound.  In this section, we consider Einstein-Maxwell-Dilaton (EMD) theories in general dimensions, with the Lagrangian
\be
{\cal L} = \sqrt{-g} \Big(R - \ft14 e^{a\phi} F^2 - \ft12 (\partial\phi)^2\Big)\,.
\ee
It is convenient to parameterize the dilaton coupling constant $a$ by
\be
a^2=\fft{4}{N} - \fft{2(D-3)}{D-2}\,.
\ee
Thus $a^2\ge 0$ implies that $0<N\le 2(D-2)/(D-3)$.  The EMD theories are supergravity inspired in that the theories of integer $N$ can be embedded in appropriate supergravities. The theories admit electrically-charged black holes, (see e.g.~\cite{Lu:2013eoa})
\bea
ds^2 &=& -H^{-\fft{D-3}{D-2} N} \tilde f dt^2 +
H^{\fft{N}{D-2}} \Big(\fft{dr^2}{\tilde f} + r^2 d\Omega^2\Big)\,,\quad
A = \psi dt\,,\quad \phi = \ft12 N a \log H\,,\nn\\
\tilde f &=& 1 - \fft{\mu}{r^{D-3}}\,,\qquad \psi = \fft{\sqrt{N\,q(\mu+q)}}{r^{D-3} H}\,,\qquad H=1 + \fft{q}{r^{D-3}}\,.
\eea
In other words, we have
\be
h=H^{-\fft{D-3}{D-2} N} \tilde f\,,\qquad
f=H^{-\fft{N}{D-2}} \tilde f\,,\qquad \rho=r^2 H^{\fft{N}{D-2}}\,.
\ee
The solutions reduce to the RN black holes when $a=0$, corresponding to $N=2(D-2)/(D-3)$.  The black hole horizon is located at $r=r_+$, given by $\mu = r_+^{D-3}$. The surface gravity on the horizon is
\be
\kappa = \fft{D-3}{2r_+} \Big(1 + \fft{q}{r_+^{D-3}}\Big)^{-\fft12 N}\,.
\ee
The black hole thermodynamical quantities can be easily obtained by standard procedure, and they are
\bea
M&=&\fft{(D-2)\Omega}{16\pi} \Big(\mu + \fft{D-3}{D-2} N q\Big)\,,\qquad
T=\fft{\kappa}{2\pi}\,,\qquad S=\ft14 \Omega\, r_+^{D-2} H(r_+)^{\fft12 N}\,,\nn\\
\Phi &=& \psi(r_+)\,,\qquad Q=\fft{(D-3)\Omega}{16\pi} \sqrt{Nq (\mu + q)}\,.
\eea
It is easy to verify that the first law $dM=TdS + \Phi dQ$ of black hole thermodynamics is satisfied.

The Lorentz force term in the particle action (\ref{particleaction}) can now be modified by a factor $e^{b\phi}$, and in particular when $b=\ft12a$ the constant shift symmetry of the dilaton of the EMD theory is maintained by the modified Lorentz force law.  However, this symmetry is not sacred since it can be easily violated by a scalar potential. Thus for simplicity, we shall take $b=0$. The equilibrium $r_0$ of a charged particle of mass and charge $(m,e)$ around the black hole is determined by
\be
\fft{e}{m} = \fft{\Omega}{32\pi} \fft{h'(r_0) H(r_0)^2}{Q\sqrt{h(r_0)}}r_0^{D-2}\,.
\ee
For the equilibria lying within $(r_+,\infty)$, the charge/mass ratio must satisfy
\be
\fft{e}{m}>\fft{D-3}{2(D-2)} \fft{M}{Q}\,.
\ee
It turns out that the equilibria are all unstable, and the Lyapunov exponent is given by
\bea
\lambda^2 &=&\fft{N^2 (D-3)^2 r_0^{2(5-2D)}}{64(D-2)^2H(r_0)^{N +2}}\Big[4(D-3)^2
H(r_0)^2 (r_0 r_+)^{2(D-3)}\nn\\
&&\qquad +4(D-2)(D-3)a^2\Big(1 + (r_0r_+)^{2(D-3)} + 2q r_0^{2(D-3)} r_+^{2(D-3)}\nn\\
&&\qquad\qquad\qquad +
q^2\big(2r_0^{D-3}(r^{D-3}_0-r_+^{D-3}) + r_+^{2(D-3)}\big)\Big)\nn\\
&&\qquad + (D-2)^2H(r_0)\,(r_0 r_+)^{D-3} \big((r_0 r_+)^{D-3} + q (4r^{D-3}_0-3
r_+^{D-3})\big)  \Big]\,,
\eea
which is positive definite for $r_0>r_+$.  For two cases, the expression $\lambda/\kappa$ becomes particularly simple:
\bea
a=0:&&\qquad \fft{\lambda}{\kappa} = \Big(\fft{q + r_+^{D+3}}{q + r_0^{D-3}}\Big)^{\fft{D-2}{D-3}}\,,\nn\\
a=1\,,\quad D=4:&&\qquad \fft{\lambda}{\kappa}=\Big(\fft{q + r_+}{q + r_0}\Big)^2\,.\label{twocases}
\eea

We now examine the chaotic motion bound (\ref{unicmb}).  For the above two cases in (\ref{twocases}) the bound is satisfied.  In general, we find that this bound can hold or be violated depending on the value of the dilaton coupling constant $a$.  We shall not classify all the possibilities here, but instead give some explicit examples.  First we consider $D=4$, and analytical results can be explicitly given for the $N=1,2,3,4$ integer values.  The $N=4$ example is precisely the RN black hole written in different coordinates. Introducing a dimensionless parameter $x$ as in (\ref{xydef}), we find that in $D=4$:
\bea
&&N=1:\nn\\
\kappa^2 - \lambda^2 &=&
\frac{x}{16 r_+ (x+1)^3 \left(q+r_+\right) \left(q+r_+ (x+1)\right){}^3}\Big[
q^3 x (4 x+9)\nn\\
&&+q^2 r_+ \left(12 x^3+48 x^2+57 x+16\right)+4 q r_+^2 (x+1) \left(3 x^3+12 x^2+18 x+8\right)
\nn\\
&&+4 r_+^3 (x+1)^2 (x+2) \left(x^2+2 x+2\right)
\Big]\,;\nn\\
&& N=2:\nn\\
\kappa^2 - \lambda^2 &=&\frac{r_+ x \left(2 q+r_+ (x+2)\right) \left(2 q^2+2 q r_+ (x+2)+r_+^2 \left(x^2+2 x+2\right)\right)}{4 \left(q+r_+\right){}^2 \left(q+r_+ (x+1)\right){}^4}\,;\nn\\
&& N=3:\nn\\
\kappa^2 - \lambda^2 &=& \frac{r_+ x}{16 (x+1) \left(q+r_+\right){}^3 \left(q+r_+ (x+1)\right){}^5}
\Big[q^4 r_+ (7 x+16)-3 q^5 x\nn\\
&&+q^3 r_+^2 \left(40 x^2+95 x+64\right)+q^2 r_+^3 \left(40 x^3+160 x^2+213 x+96\right)\nn\\
&&+4 q r_+^4 \left(5 x^4+25 x^3+50 x^2+46 x+16\right)\nn\\
&&+4 r_+^5 (x+1)^2 \left(x^3+4 x^2+6 x+4\right)
\Big]\,;\nn\\
&&N=4:\nn\\
\kappa^2 - \lambda^2 &=&\frac{r_+^3 x \left(2 q+r_+ (x+2)\right) \left(2 q^2+2 q r_+ (x+2)+r_+^2 \left(x^2+2 x+2\right)\right)}{4 \left(q+r_+\right){}^4 \left(q+r_+ x+r_+\right){}^4}\,.
\eea
It is thus clear that for $N=1,2$ and 4, we have always $\kappa^2-\lambda^2>0$ and hence the chaotic motion bound (\ref{unicmb}) is satisfied globally. On the other hand, when $N=3$, there is a negative $-3q^5 x$ term, which can cause $\kappa^2 - \lambda^2$ to be negative at certain $x$, for sufficiently large $q$ or small $r_+$. Near the horizon with small $x$, we find the Taylor expansion
\bea
\kappa^2 - \lambda^2 &=& \fft{r_+^{2+N}}{r_0^{N+6} H(r_0)^{N+2}}\Big[r_+ (r_+ +q) x\nn\\
&& \qquad +\big(\ft{3}{16} (4-N)(2-N) q^2 + 4 q r_+ + \ft12 (2N+7) r_+^2\big) x^2 + {\cal O}(x^3)\Big]\,.
\eea
Since the leading-order term is positive, it follows that the bound (\ref{unicmb}) holds locally at the near-horizon region.  For $2<N<4$, and sufficiently large $q$, the bound (\ref{unicmb}) can be violated at some $x$ away from the horizon. The bigger the parameter $q$, the smaller the local region where (\ref{unicmb}) is satisfied.

The situation is more complicated in general dimensions and we find that the bound (\ref{unicmb}) can be violated even locally near the horizon.  As a concrete example, we consider $D=5$ and $N=1$, and we find
\bea
\kappa^2-\lambda^2 &=& \frac{x (x+2)}{9 (x+1)^4 \left(q+r_+^2\right) \left(q+r_+^2 (x+1)^2\right){}^3}
\Big[q^3 \left(x^2+2 x-6\right)\nn\\
&&+q^2 r_+^2 \left(27 x^4+108 x^3+157 x^2+98 x+15\right)\nn\\
&&+3 q r_+^4 (x+1)^2 \left(9 x^4+36 x^3+63 x^2+54 x+16\right)\nn\\
&&+9 r_+^6 (x+1)^4 \left(x^4+4 x^3+7 x^2+6 x+3\right)
\Big]\,.
\eea
This quantity is not positive definite.  In particular, as $x\rightarrow 0$, we have the leading term
\be
\kappa^2-\lambda^2\Big|_{x\rightarrow 0} = -\frac{2 x \left(2 q-9 r_+^2\right)}{3 \left(q+r_+^2\right){}^2} + {\cal O}(x^2)\,.
\ee
Thus we see that the bound (\ref{unicmb}) is violated even at the near horizon region since the above leading-order term near the horizon becomes negative when $q> \ft92 r_+^2$.

For general $D$ and $N$, we have the leading order of the near-horizon expansion
\be
\kappa^2-\lambda^2\Big|_{x\rightarrow 0}=
\fft{(D-2)(D-3)^2}{2r_+^{D-1} H(r_+)^{1+N}} \Big(r_+^{D-3}  - \fft{(D-4)a^2 q}{(D-2)a^2 + 2(D-3)}\Big)x + {\cal O}(x^2)\,.
\ee
Thus we see that unless $a=0$ or $D=4$, the quantity can be negative for sufficiently large $q$.
Thus the violation of chaotic motion bound (\ref{unicmb}) is rather a common occurrence for charged black holes in EMD theories.  Note that charged black holes in EMD theories belong to the general rather that the special static black holes described in \cite{Li:2017ncu}.  Thus the general discussion for the special static black holes in section \ref{sec:setup} does not apply and indeed we found above the explicit examples where the chaotic motion bound (\ref{unicmb}) is even violated locally by equilibria that are in the vicinity of the horizon.

Owing to the fact that we have found examples of asymptotically-flat black holes in EMD theories for which the bound (\ref{unicmb}) is violated, we shall not investigate further asymptotically-(A)dS black holes in this category.

\section{Further charged black holes}
\label{sec:more}

We have seen that for asymptotically flat or AdS black holes, the chaotic motion bound (\ref{unicmb}) is satisfied by the charged particles around the RN black holes, but can be violated when around the charged black holes of EMD theories.  The RN black holes belong to the special static solutions whilst the charged black holes in EMD theories belong to the general static solutions.  It is thus of interest to investigate further examples of special static black holes and examine whether the bound can be violated.  Since we have already found asymptotically-dS examples in section \ref{sec:RNdS} that violate the bound, we shall consider here only those that are asymptotic to Minkowski or AdS spacetimes.

\subsection{Einstein-Born-Infeld Theory}
\label{sec:EBI}

In this subsection, we consider Einstein-Born-Infeld (EBI) theory and the Lagrangian is
\be
{\cal L}=\sqrt{-g} (R - 2\Lambda_0) - b^2 \sqrt{-\det (g_{\mu\nu} + \fft{F_{\mu\nu}}{b})}\,,
\ee
where the bare cosmological constant is related to the effective cosmological constant by $\Lambda_0=\Lambda - b^2/2$.  The theory reduces to Einstein-Maxwell gravity when the Born-Infeld parameter $b$ takes the $b\rightarrow \infty$ limit.

We focus our discussion in $D=4$ and the static solution is of the special kind, given by \cite{gsp,Breton:2002td,Cai:2004eh,Li:2016nll}
\bea
h &=& f = -\ft13 \Lambda_0 r^2 + k - \fft{2M}{r} - \fft{b^2}{6} \sqrt{r^4 + \fft{q^2}{b^2}} +
\fft{Q^2}{3r^2}\, {}_2F_1[\ft14,\ft12;\ft54;-\ft{q^2}{b^2 r^4}]\,,\nn\\
\psi' &=& \fft{q}{\sqrt{r^4 + \fft{q^2}{b^2}}}\,,\qquad \rho=r\,.
\eea
The large-$r$ expansion of $f$ is
\be
f= -\ft13\Lambda r^2 + k - \fft{2M}{r} + \fft{q^2}{r^2} + \cdots\,.
\ee
Furthermore, the $b\rightarrow \infty$ limit yields the RN black holes.  Since the black hole belongs to the special static solutions, it follows from the discussion in section \ref{sec:setup} that the chaotic motion bound (\ref{unicmb}) is satisfied in the vicinity of the horizon, since we have
\be
\fft{\psi''}{\psi'} <0\,.
\ee
In this subsection, we examine whether the bound (\ref{unicmb}) holds globally or not for finite $b$.

\subsubsection{Asymptotically flat}

Here we set $\Lambda=0$, in which case, we must choose $k=1$. For sufficiently large $M$, there exists a horizon $r_+>0$ satisfying $f(r_+)=0$.  The corresponding surface gravity is
\be
\kappa =\fft{1}{2r_+} + \fft{b^2}{4r_+} \Big( r_+^2 - \sqrt{r_+^4 + \fft{Q^2}{b^2}}\,\Big)\,.
\ee
The condition $\kappa\ge 0$ implies that $Q$ has an upper limit for fixed $r_+$
\be
Q\le Q_{\rm ext} \equiv 2\sqrt{r_+^2 + \ft{1}{b^2}}\,.
\ee
The extremal solution corresponds to $Q=Q_{\rm ext}$ for which we have $\kappa=0$. For $Q<Q_{\rm ext}$, there exist an inner horizon $r_-$, which does not concern us in our discussion here.

The equilibrium $r_0$ and the corresponding Lyapunov exponent $\lambda$ are given by (\ref{emgen}) and (\ref{lambdagen}) respectively.  We shall not present the explicit results owing to the complexity of the formulae, but instead give some qualitative descriptions of the properties.

We first consider the extremal solution $Q=Q_{\rm ext}$ for which we have $\kappa=0$.  In this case, the existence of any unstable equilibrium implies the violation of the chaotic motion bound (\ref{unicmb}). In the limit of $b\rightarrow \infty$, we have $\lambda=0$, leading to the no-force condition described in section \ref{sec:RNflat} for the asymptotically-flat extremal RN black holes.  In the large $b$ expansion, we find
\be
\lambda^2 = \frac{\left(r_0-r_+\right){}^3 \left(30 r_+^3-6 r_0 r_+^2-3 r_0^2 r_+-r_0^3\right)}{5 r_0^{10} b^2} + {\cal O}(\fft{1}{b^4})\,.
\ee
Thus we see that for the large but finite $b$, the leading-order term has positive $\lambda^2$ in $0<r_0<r_0^*$ where $r_0^*=1.977 r_+$ and hence the equilibria are unstable.  Equilibria located at $r_0>r_0^*$ are all stable.  This feature turns out to be true for all $b$.  Near the horizon, we find
\be
\lambda^2 = \fft{4b^4(2 + 3 b^2 r_+^2)}{3r_+ (2 + b^2 r_+^2)^4} (r_0-r_+)^3 + {\cal O}((r_0-r_+)^4)\,,
\ee
which is positive.  In the large $r_0$ expansion, we have
\bea
\lambda^2 &=& \fft{c}{r_0^4} + {\cal O}(r_0^{-6})\,,\nn\\
c &=& \frac{1}{9 b^4 r_+^2}\Big[-b^2 r_+^2 \left(8 b^2 r_+^2+9\right)+4 \left(b^2 r_+^2+1\right)^2 \, _2F_1[\ft{1}{4},\ft{1}{2};\ft{5}{4};-\ft{4 \left(b^2 r_+^2+1\right)}{b^4 r_+^4}]^2\nn\\
&&+4 \left(b^4 r_+^4+b^2 r_+^2\right) \, _2F_1[\ft{1}{4},\ft{1}{2};\ft{5}{4};-\ft{4 \left(b^2 r_+^2+1\right)}{b^4 r_+^4}]\Big]\,.
\eea
It can be easily demonstrated numerically that the coefficient $c$ is negative for all $b$ and $r_+>0$.
Thus we see that $\lambda^2=0^+$ as $r_0\rightarrow (r_+)^+$ and as $r_0$ increases, $\lambda^2$ reaches a maximal and then reduces to zero at certain $r_0^*>r_+$, after which $\lambda^2$ becomes negative and the equilibria become all stable.  This is different from asymptotically-flat extremal RN black holes, where the equilibria are all marginally stable.  The fact that unstable equilibria exist in the extremal black hole with $\kappa=0$ implies that the bound (\ref{unicmb}) is maximally violated.

\begin{figure}[ht]
\begin{center}
\ \ \ \ \ \includegraphics[width=7cm]{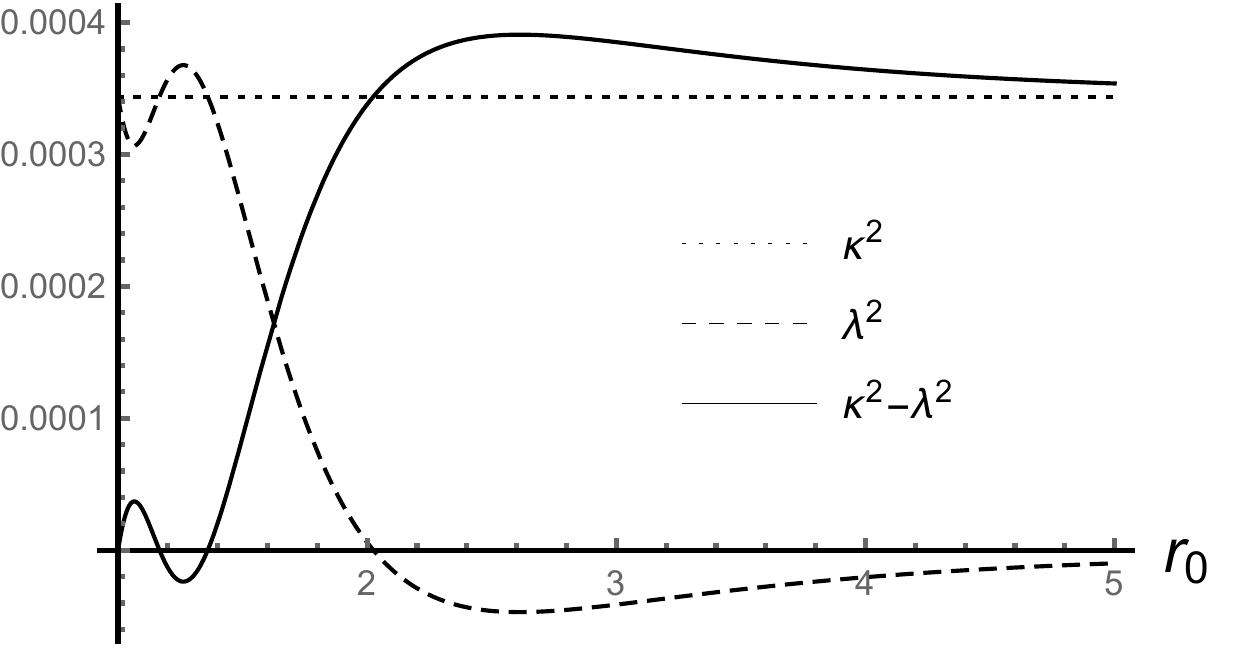}
\caption{An explicit example of violation of the bound (\ref{unicmb}) by a non-extremal black hole in EBI theory. The black hole parameters are given by $r_+=1$, $b=5$ and $Q=2$.  It can be seen that the bound is satisfied at the vicinity of the black hole horizon $r_0\in (1,1.2)$, but violated in region $(0.2,0.4)$, satisfied again in (0.4,2). The $r_0>2$ equilibria become stable.}
\label{BIF-flat-1}
\end{center}
\end{figure}

We now consider non-extremal solutions with $\kappa >0$, achieved by setting $Q<Q_{\rm ext}$.  As a concrete example, we set $Q=2 r_+$, which becomes extremal when $b\rightarrow \infty$.  For any finite $b$, we have $Q<Q_{\rm ext}$.
As the equilibrium $r_0$ approach the horizon $r_+$, we $\lambda \rightarrow \kappa$ and hence the equilibrium is unstable.  We find that
\be
\kappa^2 - \lambda^2 =\frac{2 b^2 \left(r_0-r_+\right)}{r_+ \left(b^2 r_+^2+4\right) \left(b^4 r_+^4+4 b^2 r_+^2+b r_+ \sqrt{b^2 r_+^2+4} \left(b^2 r_+^2+2\right)+2\right)} + {\cal O}((r-r_+)^2)\,.
\ee
For sufficiently small $b$, the chaotic motion bound (\ref{unicmb}) can be globally satisfied by all equilibria.  When $b$ increases, the situation becomes more complicated, and we present the behavior of $(\kappa^2,\lambda^2)$ and $(\kappa^2-\lambda^2)$ in Fig.~\ref{BIF-flat-1}.  It is worth pointing out that in this $Q=2r_+$ case, the ratio $\lambda/\kappa$ can be arbitrarily large at certain unstable equilibria for sufficiently large $b$. The surface gravity approaches zero in this limit.

When $Q$ is even smaller than $2r_+$, such that the black hole is non-extremal for all $b$, we find that the chaotic motion bound (\ref{unicmb}) is satisfied globally.

\subsubsection{Asymptotically AdS}

We now take the bare cosmological constant to be
\be
\Lambda_0=-\ft12 b^2 - \fft{3}{\ell^2}\,.
\ee
The vacuum solution is the AdS with radius $\ell$.  We shall consider only the $k=1$ case as a representative solution and hence we have
\be
\kappa = \fft{3 r_+}{2\ell^2} + \fft{1}{2r_+} + \fft{b^2}{4r_+} \Big( r_+^2 - \sqrt{r_+^4 + \fft{Q^2}{b^2}}\,\Big)\,.
\ee
The extremal solution, corresponding to $\kappa=0$, can arise if we take $Q=Q_{\rm ext}$ with
\be
Q_{\rm ext} = \ft{2}{b}\sqrt{\big(1 + \ft{3r_+^2}{\ell^2}\big)\big(1 + b r_+^2 + \ft{3r_+^2}{\ell^2}\big)}\,.
\ee
In this extremal limit, we find, after setting $\ell=1$, that in the vicinity of the horizon,
\bea
\lambda^2 &=& \frac{4 \left(b^2 \left(6 r_+^2+1\right)+18 r_+^2+6\right)}{3 r_+ \left(\left(b^2+6\right) r_+^2+2\right){}^4} \Big(-3 b^6 r_+^6+3 b^4 \left(3 r_+^3+r_+\right)^2+12 \left(3 r_+^2+1\right)^3\nn\\
&&\qquad+2 \left(12 r_+^2+1\right) \left(3 b r_+^2+b\right)^2\Big) (r_0-r_+)^3 + {\cal O}((r_0-r_+)^4)\,.
\eea
Thus we see that for a given $r_+$, $\lambda^2$ is negative for large $b$, and hence the equilibria are stable in the vicinity of the horizon.  It is intriguing that charged particles can be trapped in stable equilibria near the horizon.  As $b$ decreases, $\lambda^2$ becomes positive and the equilibria become unstable.  Since $\kappa=0$, the chaotic motion bound (\ref{unicmb}) is violated even at the vicinity of the horizon.  Far away from the horizon, $\lambda^2$ can become negative. In particular, in the large $r_0$ expansion, we find that
\be
\lambda^2 = - \fft{2r_0^2}{\ell^4} - \fft{3}{\ell^2} + {\cal O}(r_0^{-1})\,.
\ee
Thus for sufficiently large $r_0$, the equilibria are all stable regardless the values of $b$.

For non-extremal black holes, we find that the characteristics is similar to that of RN-AdS black holes:
$\lambda^2$ is positive for equilibria close to the horizon, but becomes negative away from the horizon, with the bound (\ref{unicmb}) always satisfied.

\subsection{Einstein-Gauss-Bonnet-Maxwell gravity}

As the final example of this paper, we now consider Einstein-Gauss-Bonnet-Maxwell (EGBM) theory:
\be
{\cal L}=\sqrt{-g} \Big(R - 2\Lambda_0 + \alpha_{\rm GB} (R^{\mu\nu\rho\sigma} R_{\mu\nu\rho\sigma} -
4R^{\mu\nu}R_{\mu\nu} +R^2) - \ft14 F^2\Big)\,,
\ee
The theory admits a special static charged black hole \cite{Wiltshire:1985us,Cvetic:2001bk}, given by
\bea
h=f &=& k + \fft{r^2}{2\alpha} \Big(1 - \beta \sqrt{1 + \ft{8\alpha\,M}{\beta^2 r^{D-1}} -
\ft{4\alpha q^2}{\beta^2r^{2(D-2)}}}\,\Big)\,,\nn\\
\psi &=& \sqrt{\ft{2(D-2)}{D-3}} \fft{q}{r^{D-3}}\,,\qquad \beta = 1 - \fft{2\alpha}{\ell^2}\,,\nn\\
\Lambda_0 &=& \ft12 (D-1)(D-2) \big(\fft{1}{\ell^2}-\fft{\alpha}{\ell^4}\big)\,,\qquad
\alpha=\fft{\alpha_{\rm GB}}{(D-3)(D-4)}\,.
\eea
(The chargeless solutions can be found in \cite{Boulware:1985wk,Cai:2001dz}.) It is easy to verify that in the $\alpha\rightarrow 0$ limit, the solution reduces to the general RN solution (\ref{genrnsol}). For simplicity, we shall discuss the $D=5$ example only.  We also assume that $\alpha>0$, in which case, the absence of ghosts implies that $\beta> 0$.  The $\beta=0$ case yields gravity without linear gravitons \cite{Fan:2016zfs}.

\subsubsection{Asymptotically flat}

In this case, we set $\ell\rightarrow \infty$ and $k=1$.  The solution in general have two horizons, in terms of which, the mass and charge parameters can be expressed as
\be
M=\ft12 (r_-^2 + r_+^2 + \alpha)\,,\qquad q = r_- r_+\,.
\ee
The surface gravity on the horizon is
\be
\kappa = \fft{r_+^2 - r_-^2}{r_+(r_+^2 + 2 \alpha)}\,.
\ee
In the extremal limit $r_-=r_+$, we have
\bea
r_0\rightarrow r_+:&& \lambda^2 = -\fft{64\alpha}{r_+ (2\alpha + r_+^2)^3} (r_0-r_+)^3 + {\cal O}((r_0-r_+)^4)\,,\nn\\
r_0\rightarrow \infty:&& \lambda^2 =\fft{\alpha(4r_+^2 +\alpha)}{r_0^6} + {\cal O}((r_0^{-8})\,.
\eea
With our assumption $\alpha>0$, we see that the equilibria are stable near the horizon, and become unstable distance away from the horizon, causing the violation of the bound (\ref{unicmb}).

For non-extremal, the general discussion in section \ref{sec:setup} applies and we we have
\be
\kappa^2 -\lambda^2 =\fft{6\kappa^2}{r_+} (r_0-r_+)\,,
\ee
For higher enough surface gravity, the chaotic bound (\ref{unicmb}) is globally satisfied by all
equilibria; for sufficiently small $\kappa$, the chaotic bound (\ref{unicmb}) can be violated outside of the horizon.  The ratio $\lambda/\kappa$ can be arbitrary large for sufficiently small $\kappa$. As a concrete example, we consider $(\alpha, r_-, r_+)= (1,99/100,1)$, the chaotic bound is violated at $r_0\in (4.3,6.2)$.

\subsubsection{Asymptotically AdS}

We now consider the case with finite AdS radius $\ell$.  The black hole solution in general have two horizons and we can express the mass and charge parameters as
\bea
M &=& \ft12 (r_-^2 + r_+^2 + \alpha) + \big(\fft1{2\ell^2} - \fft{\alpha}{2\ell^4}\big)
(r_-^4 + r_-^2 r_+^2 + r_+^4)\,,\nn\\
q&=& r_- r_+ \sqrt{1 + \big(\ft{1}{\ell^2} -\ft{\alpha}{\ell^4}\big) (r_-^2 + r_+^2)}\,.
\eea
The surface gravity on the horizon is
\be
\kappa = \fft{r_+^2-r_-^2}{r_+ (r_+^2 + 2\alpha)} \Big(1 + \fft{\ell^2-\alpha}{\ell^4}
(r_-^2 + 2 r_+^2)\Big)\,.
\ee
In the extremal limit ($r_-=r_+$), we have, near the horizon, that
\bea
\lambda^2 &=& - \fft{8 (r_0-r_+)^3}{\ell^4 r_+ (r_+^2 + 2\alpha)^3}
(2\ell^2 + 3 r_+^2 + 3 \beta r_+^2)\Big( 2(1-\beta) \ell^4 + 4 (1-\beta^2) \ell^2 r_+^2 +
(1+\beta) r_+^4\Big)\cr
&& + {\cal O}{(r_0 - r_+)^4}\,.
\eea
At the asymptotic infinity, on the other hand, we have $\lambda^2 = - 3r_0^2/\ell^4 + {\cal O}(1)$.  For sufficiently small $\ell$, we find that $\lambda^2$ is negative for all equilibria outside the horizon.  However, for sufficiently large $\ell$, the situation becomes the same as that in the asymptotically-flat case where the $\lambda^2$ can be positive in some regions outside of the horizon, thus violate the bound (\ref{unicmb}).

For non-extremal black holes, the equilibria near the horizon is unstable, and satisfying the bound (\ref{unicmb}), consequence of being special static black holes, as discussed in section \ref{sec:setup}. The equilibria become stable away from the horizon.  For small enough $\ell$, the region of stable equilibria extends all the way to asymptotic infinity. For large $\ell$, new unstable region can emerge in some local region outside and away from horizon and the chaotic motion bound (\ref{unicmb}) can be violated in these unstable regions.

\section{Conclusions}
\label{sec:con}

Charged particles around a static charged black hole can have a static equilibrium sphere (or some hypersurface) outside the horizon $r_+$. The radius $r_0$ of the equilibrium surface is determined by the charge/mass ratio of the charged particle.  The equilibria can be stable, marginal or unstable, depending on the specific property of the black hole.  It was well known that for the unstable equilibrium, the general perturbative motion of the particle in the equilibrium sphere is chaotic.  An intriguing question is whether there is a universal upper bound for the Lyapunov exponent for such chaotic motion. A natural candidate (\ref{unicmb}) was proposed in \cite{Hashimoto:2016dfz} by examining the equilibrium properties near the horizon.

In this paper, we investigated a large class of charged black holes and found that $\lambda$ approaches $\kappa$ universally for the unstable equilibria that are near the horizon. The results are consistent with those in \cite{Hashimoto:2016dfz}. The bound (\ref{unicmb}), however, is not universally satisfied, albeit it can be held by some specific black holes.  For RN and RN-AdS black holes, we found that the bound (\ref{unicmb}) is indeed globally satisfied by all the unstable equilibrium hypersurfaces.  It can be violated, however, by some unstable equilibria at some appropriate distance away from the horizon of the RN-dS black holes.

We also showed that for special static black holes (with $h=f, \rho=r$), with minimally-coupled Maxwell field, the bound (\ref{unicmb}) is universally satisfied at the local near-horizon region. However, for general static black holes, such as those in EMD theories, even this local bound can be violated in the vicinity of the horizons.

We found that unstable equilibria could arise not only in the vicinity of the black hole horizon, but also at some distance away from the black hole.  This implies that chaotic motion can emerge for a particle under the influence of a massive object, not necessarily a black hole. For the asymptotically-dS black hole, there exists an unstable equilibrium for neutral massive particles.  In particular, we find that for (spherical) mass $M$, the equilibrium is located at
\be
r_0 = \Big(\fft{3M}{\Lambda_0}\Big)^{\fft13}\,.
\ee
For sufficiently small cosmological constant $\Lambda_0$, the equilibrium $r_0$ can be large distance away from the horizon. We found that this equilibrium is always unstable, and the corresponding Lyapunov exponent (\ref{schds}) satisfies the chaos bound.  Applying this result to our galaxy, whose effective Schwarzschild radius is about 0.2 light year, the unstable equilibria is located at 2.7 million light years away, with Lyapunov exponent $\lambda=0.004s^{-1}$.

Since the chaos we studied in this paper are for a single particle outside the horizon, the violation of the chaos bound (\ref{unicmb}) is not necessarily a contradiction to this same chaos bound but for quantum thermal systems with a large number degrees of freedom.  Nevertheless, it raises an intriguing question how the bound is restored with the increasing number of degrees of freedom. We have also seen explicit examples where unstable equilibria exist for extremal black holes with $\kappa=0$, the ratio $\lambda/\kappa$ can thus be arbitrarily large for sufficiently small $\kappa$.  This leads to a tantalizing question whether there is a universal upper bound of chaos beyond (\ref{unicmb}) for a single particle outside a black hole horizon. From the black holes we studied in this paper, we find that for sufficiently large $\kappa$, there may indeed be a universal upper bound
\be
\fft{\lambda}{\kappa} < {\cal C}\,,
\ee
where ${\cal C}$ is some order-one constant.

\section*{Acknowledgement}

The work is supported in part by NSCF grants No.~11475024 and No.~11875200.


\begin{thebibliography}{99}

\bibitem{Sota:1995ms}
Y.~Sota, S.~Suzuki and K.i.~Maeda,
``Chaos in static axisymmetric space-times. 1: Vacuum case,''
Class.\ Quant.\ Grav.\  {\bf 13} (1996) 1241
doi:10.1088/0264-9381/13/5/034
[gr-qc/9505036].

\bibitem{Suzuki:1996gm}
S.~Suzuki and K.i.~Maeda,
``Chaos in Schwarzschild space-time: The motion of a spinning particle,''
Phys.\ Rev.\ D {\bf 55} (1997) 4848
doi:10.1103/PhysRevD.55.4848
[gr-qc/9604020].

\bibitem{Cardoso:2008bp}
V.~Cardoso, A.S.~Miranda, E.~Berti, H.~Witek and V.T.~Zanchin,
``Geodesic stability, Lyapunov exponents and quasinormal modes,''
Phys.\ Rev.\ D {\bf 79} (2009) 064016
doi:10.1103/PhysRevD.79.064016
[arXiv:0812.1806 [hep-th]].

\bibitem{Dalui:2018qqv}
S.~Dalui, B.~Ranjan Majhi and P.~Mishra,
``Presence of horizon makes particle motion chaotic,''
arXiv:1803.06527 [gr-qc].

\bibitem{Hashimoto:2016dfz}
K.~Hashimoto and N.~Tanahashi,
``Universality in chaos of particle motion near black hole horizon,''
Phys.\ Rev.\ D {\bf 95} (2017) no.2,  024007
doi:10.1103/PhysRevD.95.024007
[arXiv:1610.06070 [hep-th]].

\bibitem{Maldacena:2015waa}
  J.~Maldacena, S.H.~Shenker and D.~Stanford,
  ``A bound on chaos,''
  JHEP {\bf 1608}, 106 (2016)
  doi:10.1007/JHEP08(2016)106
  [arXiv:1503.01409 [hep-th]].

\bibitem{Li:2017ncu}
  Y.Z.~Li, H.S.~Liu and H.~L\"u,
  ``Quasi-topological Ricci polynomial gravities,''
  JHEP {\bf 1802}, 166 (2018)
  doi:10.1007/JHEP02(2018)166
  [arXiv:1708.07198 [hep-th]].

\bibitem{Lu:2013eoa}
  H.~L\"u,
  ``Charged dilatonic ads black holes and magnetic AdS$_{D-2} \times R^{2}$ vacua,''
  JHEP {\bf 1309}, 112 (2013)
  doi:10.1007/JHEP09(2013)112
  [arXiv:1306.2386 [hep-th]].

\bibitem{gsp} A.~Garc\'ia, H.~Salazar and J.F.~Pleb\'anski, ``Type-D solutions of the Einstein and Born-Infeld nonlinear-electrodynamics equations,'' Nuovo Cimento B 84, 65 (1984).

\bibitem{Breton:2002td}
  N.~Breton,
 ``Geodesic structure of the Born-Infeld black hole,''
  Class.\ Quant.\ Grav.\  {\bf 19}, 601 (2002).
  doi:10.1088/0264-9381/19/4/301

\bibitem{Cai:2004eh}
  R.G.~Cai, D.W.~Pang and A.~Wang,
  ``Born-Infeld black holes in (A)dS spaces,''
  Phys.\ Rev.\ D {\bf 70}, 124034 (2004)
  doi:10.1103/PhysRevD.70.124034
  [hep-th/0410158].

\bibitem{Li:2016nll}
  S.~Li, H.~L\"u and H.~Wei,
  ``Dyonic (A)dS black holes in Einstein-Born-Infeld theory in diverse dimensions,''
  JHEP {\bf 1607}, 004 (2016)
  doi:10.1007/JHEP07(2016)004
  [arXiv:1606.02733 [hep-th]].

\bibitem{Wiltshire:1985us}
  D.L.~Wiltshire,
 ``Spherically symmetric solutions of Einstein-maxwell theory with a {Gauss-Bonnet} term,''
  Phys.\ Lett.\  {\bf 169B}, 36 (1986).
  doi:10.1016/0370-2693(86)90681-7

\bibitem{Cvetic:2001bk}
  M.~Cveti\v c, S.~Nojiri and S.D.~Odintsov,
``Black hole thermodynamics and negative entropy in de Sitter and anti-de Sitter Einstein-Gauss-Bonnet gravity,'' Nucl.\ Phys.\ B {\bf 628}, 295 (2002)
  doi:10.1016/S0550-3213(02)00075-5
  [hep-th/0112045].

\bibitem{Boulware:1985wk}
  D.G.~Boulware and S.~Deser,
{\it String generated gravity models,}
  Phys.\ Rev.\ Lett.\  {\bf 55}, 2656 (1985).
  doi:10.1103/PhysRevLett.55.2656

\bibitem{Cai:2001dz}
  R.G.~Cai,
{\it Gauss-Bonnet black holes in AdS spaces,}
  Phys.\ Rev.\ D {\bf 65}, 084014 (2002)
  doi:10.1103/PhysRevD.65.084014
  [hep-th/0109133].

\bibitem{Fan:2016zfs}
  Z.Y.~Fan, B.~Chen and H.~L\"u,
  ``Criticality in Einstein-Gauss-Bonnet gravity: gravity without graviton,''
  Eur.\ Phys.\ J.\ C {\bf 76}, no. 10, 542 (2016)
  doi:10.1140/epjc/s10052-016-4389-x
  [arXiv:1606.02728 [hep-th]].


\end{thebibliography}
\end{document}